\PassOptionsToPackage{fleqn}{amsmath}
\documentclass[twocolumn,showpacs,preprintnumbers,amsmath,amssymb]{revtex4}
\usepackage{graphicx}
\usepackage{dcolumn}
\usepackage{bm}
\usepackage{amssymb}
\usepackage{amsmath}

\begin{document}
\title{Magnetic critical properties and basal-plane anisotropy of Sr$_2$IrO$_4$}

\author{L. Fruchter, D. Colson, V. Brouet}%
\affiliation{Laboratoire de Physique des Solides, C.N.R.S. UMR 8502, Universit\'{e} Paris-Sud, 91405 Orsay, France}
\affiliation{$^+$Service de Physique de l'Etat Condens\'{e}, CEA-Saclay, 91191 Gif-sur-Yvette, France}
\date{Received: date / Revised version: date}

\begin{abstract}{The anisotropic magnetic properties of Sr$_2$IrO$_4$ are investigated, using longitudinal and torque magnetometry. The critical scaling across $T_c$ of the longitudinal magnetization is the one expected for the 2D XY universality class. Modeling the torque for a magnetic field in the basal-plane, and taking into account all in-plane and out-of-plane magnetic couplings, we derive the effective 4-fold anisotropy $K_4 \approx$ 1 10$^5$ erg mole$^{-1}$. Although larger than for the cuprates, it is found too small to account for a significant departure from the isotropic 2D XY model. The in-plane torque also allows us to put an upper bound for the anisotropy of a field-induced shift of the antiferromagnetic ordering temperature.}
\end{abstract}
\pacs{71.70.Ej,75.30.Kz,75.47.Lx,75.30.Gw} 

\maketitle

\section*{\label{intro}Introduction}

The Ruddlesden-Popper series, R$_{n+1}$Ir$_n$O$_{3n+1}$ where R= Sr, Ba and n = 1,2,$\infty$, has emerged as a new playground for the study of electron correlation effects. In these compounds, while extended 5$\textit{d}$ orbitals tend to reduce the electron-electron interaction, as compared to the 3$\textit{d}$ transition metal compounds as cuprates, the strong spin orbit coupling (SOC) associated to the heavy Ir and the on-site Coulomb interaction compete with electronic bandwidth to restore such correlations\cite{Kim2008}. Sr$_2$IrO$_4$, a perovskite where a IrO$_2$ layer alternates with two SrO layer, is structurally similar to the first discovered cuprate superconductor, (La,Ba)$_2$CuO$_4$. It was early proposed that the strong SOC allows for an effective localized state, entangling spin and orbital degrees of freedom, with total angular momentum J$_{eff}$ = 1/2. This spin-orbital insulating state was proposed to be the analog of the Mott insulating state found in cuprates\cite{Kim2008}.

Sr$_2$IrO$_4$ orders antiferromagnetically below $T_c$ $\simeq$ 240~K\cite{Kim2009,Ye13}: the moments lay in the IrO$_2$ plane and, as the loss of the inversion symmetry in the non cubic structure -- due to a rotation of the oxygen octahedra -- allows for a Dzyaloshinskii-Moriya interaction, a canting of the spins ($\approx$ 9 deg.) and a ferromagnetic component occur\cite{Crawford94,Jackeli09} (Fig.~\ref{sketchAF}). The in-plane net moments are coupled in an 'up-up-down-down' way from plane to plane in zero field, and align ferromagnetically with an in-plane field $H \approx$ 0.2~T \cite{Cao98}. The initial proposition in Ref.~\cite{Jackeli09} that the pseudospin Hamiltonian may be mapped onto a simple Heisenberg Hamiltonian for a square lattice antiferromagnet received several supports \cite{Kim12,Katukuri12,Fujiyama12,Boseggia13}. Recently, however, critical magnetic fluctuations were investigated using X-ray resonant magnetic scattering above $T_c$, and were found consistent with the 2D XY model rather than with the isotropic model. Moreover, it was proposed that the basal-plane anisotropy accounts for the deviation of the critical exponent of the coherence length from the one of this model\cite{Vale15}. 

The magnetic ordering of a layered compound as Sr$_2$IrO$_4$ relies, however, on the finite transverse coupling between 2D fluctuating spins, and one cannot disregard the 3D nature of this coupling, when the ordered state is considered. So, it is necessary to also investigate the dimensionality of the fluctuations as the ordering temperature is crossed. The critical scaling of the magnetization allows to do so, as shown in section \ref{LongMag}. Besides these conventional magnetization studies, the transverse magnetization provided by torque measurements is a direct way to evaluate the additional anisotropy in the basal-plane. Section \ref{TransMag} presents such measurements, and models the system in an in-plane magnetic field to obtain an estimate of the four-fold magneto-crystalline anisotropy. It is discussed whether the measured anisotropy is able to reduce the dimensionality of the magnetic system, as proposed in Ref.~\onlinecite{Vale15}.

\section{\label{LongMag}Longitudinal magnetization}

The longitudinal magnetization of a single crystal with dimensions 1200 x 400 x \textit{c} = 120 $\mu$m$^3$ was measured in magnetic fields up to 7 T. It was grown using a self-flux technique in platinum crucibles, similar to the one in Ref.~\cite{Kim2009}. In a mean-field approach, the Weiss-molecular theory allows to predict an asymptotic linear relationship between the squared magnetization, $M(T,H)^2$ and the inverse susceptibility, $H/M$, in the vicinity of $T_c$, which is the basis for the determination of $T_c$ from the so-called Arrott plot, which displays $M^2$ vs $(H/M)$\cite{Arrott57}. Below $T_c$, such a plot may be linearly extrapolated to the positive saturation magnetization, $M_s$, while, above $T_c$, the isotherms extrapolate to negative values and intercept the $(H/M)$ axis at the inverse susceptibility $\chi_0^{-1}$; the isotherm at $T = T_c$ is the one extrapolating to the origin.

In the general case where the mean-field approach fails, a modified Arrott plot must be built, which incorporates the general scaling relations for the magnetization and susceptibility at $T_c$:

\begin{equation}
(M(T,H)/M_0)^{1/\beta} = (\chi_0 H/M)^{1/\gamma} + (T-T_c)
\label{arrott}
\end{equation}

The mean-field result is retrieved, taking $\beta$ = 1/2 and $\gamma$ = 1. In practice, the appropriate exponents are rarely easily obtained in this way, as the isotherms may be only asymptotically linear, and several sets of $\beta$ and $\gamma$ values may provide equally satisfying plots (see for instance Ref.~\onlinecite{Zhou08}). So, an unbiased procedure is desirable, and we achieved this in the following way.

\begin{figure}
\resizebox{0.95\columnwidth}{!}{%
  \includegraphics{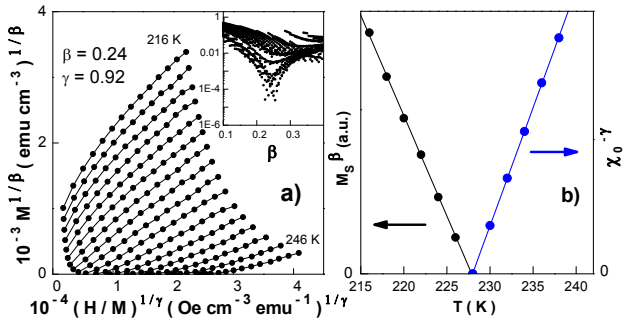}
  }
\caption{(a) Modified Arrott plot for magnetic field applied along \textit{a/b} axis. The inset displays the distance between the seed ($\beta$,$\gamma$) of the procedure and its output. (b) Asymptotic spontaneaous magnetization ($M_S$) and initial inverse susceptibility ($\chi_0^{-1}$) as obtained from (a). Lines are best linear fits of $M_s^{\beta}$ and $(\chi_0^{-1})^{\gamma}$.}\label{Arrott}
\end{figure}

\begin{figure}
\resizebox{0.9\columnwidth}{!}{%
  \includegraphics{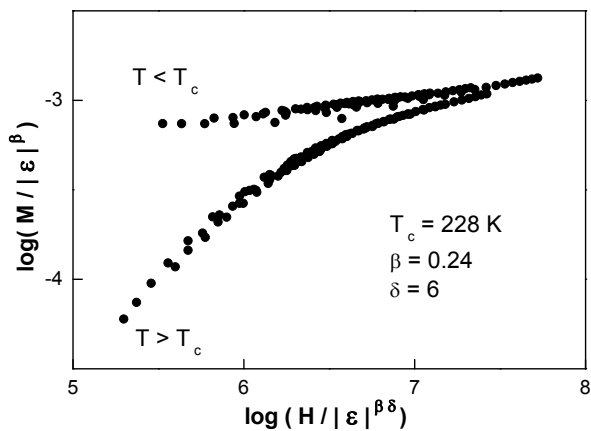}
  }
\caption{A scaling plot for $M(H)$, using $\beta$ as found in Fig.~\ref{Arrott}b.}\label{Scale}
\end{figure}
From magnetization isotherms obtained in the interval 216 K $< T <$ 246 K and 0.5 T $< H <$ 7 T, we have built the modified Arrott plot for a grid of $\beta$ and $\gamma$ values. For each of these plots, we have first determined the critical isotherm (as the one closest to a line crossing the origin); obtained the extrapolated values for $M_s$ and $\chi_0^{-1}$ (as the isotherms are not completely saturated at the maximum field in Fig.~\ref{Arrott}, we have assumed that they exponentially reach the critical isotherm slope), and finally computed the $\beta$ and $\gamma$ values from power law fits of these quantities (as in Fig.~\ref{Arrott}b). The self-consistent plot, for which the computed $\beta$ and $\gamma$ values were closest to the initial seed (the distance between the seed of the procedure and its output is shown in the inset of Fig.~\ref{Arrott}a), is obtained for $\beta$ = 0.24 $\pm$ 0.02, $\gamma$ = 0.92 $\pm$ 0.1, and $T_c$ = 228 $\pm$ 1 K.

Another way to obtain the scaling exponents is to use the reduced equation of states:

\begin{equation}
M/\left|\epsilon\right|^\beta = f_\pm(H/M^{\beta \delta})
\label{eqstate}
\end{equation}

where $f_\pm$ refer to data for $T > T_c$ and $T < T_c$ respectively, and $\epsilon$ is the reduced temperature. Using the previous values  for $\beta$ and $T_c$, the best scaling is obtained for $\delta \simeq$ 6 (Fig.~\ref{Scale}). This is compatible with the Widom scaling relation, $\delta = 1+\gamma/\beta$ = 5.2 $\pm$ 0.9. 

\begin{figure}
\resizebox{0.9\columnwidth}{!}{%
  \includegraphics{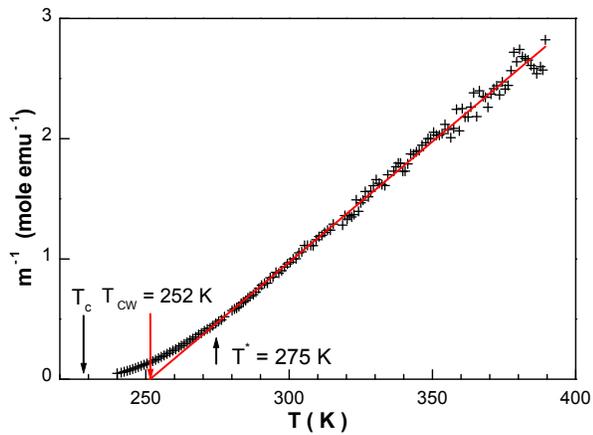}
  }
\caption{Curie-Weiss law, defining $T_{CW}$ and $T^*$ ($H$ = 0.2 T).}
\label{Curie}
\end{figure}

\begin{figure}
\resizebox{0.9\columnwidth}{!}{%
  \includegraphics{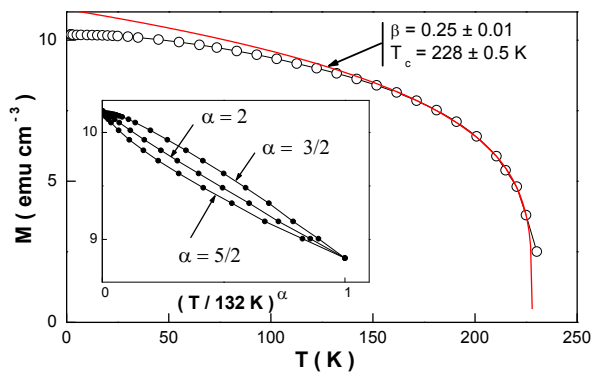}
  }
\caption{Low field magnetization data ($H$ = 0.4 T). The line is the best fit to a power-law $(T_c-T)^\beta$ for $T > 162 K$. The inset shows that $M(T \rightarrow 0$) is best described by the 3D spin 1/2 scaling, $T^2$ (a small Curie term is present -- $\approx$ 0.4\% at 1 K -- which was not removed).}
\label{scale04T}
\end{figure}

Away from the critical region, the data is well described by the conventional Curie-Weiss law, using $T_{CW}$ = 252 $\pm$ 2 K. As seen in Fig.~\ref{Curie}, the Curie-Weiss law breakdowns at a temperature $T^* \simeq$ 275 K. This is the temperature at which an atomistic description fails.
We may estimate the Levanyuk-Ginzburg criterion from the temperature range of these critical fluctuations as $t_G \simeq (T_c-T^*)/T_c \simeq 0.2$. Using the expression in 2D, $t_G = \Delta C_p^{-1} \chi_0^{-2}$, and $C_p \simeq 4 $mJ/mole K (Ref.~\onlinecite{Chikara09}), yields $\xi_0/a \simeq 10^2$ for the zero-temperature coherence length. Remarkably, the applied magnetic field does not seem to change the dimensionality of the critical fluctuations towards a 1D behavior, as is often observed (see Ref.~\cite{Kobler10} for a review). This is also evidenced by low field magnetization as in Fig.~\ref{scale04T}, for which a power-law fit yields $\beta = 0.25 \pm 0.01$, in agreement with the high-field scaling analysis. Surprisingly, the scaling for $T \rightarrow 0$ indicates neither a spin 1/2, 2D ($T^{3/2}$), nor a 2D anisotropic or 1D universal behavior ($T^{5/2}$), but clearly a 3D one ($T^2$ - Fig.~\ref{scale04T}). This could be an indication that the transverse coupling cannot be neglected in the weakly fluctuating regime.

These results indicate clearly that the magnetic transition is dominated by critical fluctuations, which are not in the mean-field universality class. The value found for $\beta$ is compatible with the 2D XY model ($\beta$ = 0.23), but are hardly compatible with a strong in-plane anisotropy, for which the exponent is pushed toward the one of the 2D Ising model ($\beta$ = 0.125), as discussed in Ref.~\onlinecite{Vale15}. In the following, we investigate the in-plane anisotropic properties, using torque magnetometry.

\section{\label{TransMag}Torque measurements}

\subsection{Experimental results}

Torque magnetometry essentially measures the magnetization component transverse to the applied field, being $\Gamma = \textbf{M} \times \textbf{B}$. Thus, for a magnetic field applied in the easy \textit{a}-\textit{b} plane, it senses the deviation of the magnetization direction from the applied field one, as a result of the basal-plane magnetocrystalline anisotropy, which tends to align the spins with specific directions.

Torque was measured using a home-made setup, built from AFM piezolevers\cite{Zech96}. This very sensitive device cannot accommodate large samples, and a smaller crystal was used, selected from the same batch as for the crystal used for conventional magnetometry. This parallepipedic sample (240 x 240 x \textit{c} = 100 $\mu m^3$) had a $T_c$ identical to the one of the larger crystal, and we could also check that its anisotropic magnetic properties (characterized by the characteristic field $H_2$ -- see below) were also identical. Torque measurement were performed by rotating the magnetic field in the \textit{a}-\textit{b} plane, in magnetic fields up to 9 T. Torque signals showed a small two-fold component -- typically 10\% of the four-fold component -- which we assign to a small misalignment of the rotation plane from the \textit{a}-\textit{b} one, so that torque also picks up part of the axial strong anisotropy. This component was systematically subtracted from the torque data as a function of the magnetic field angle, as was done for the one displayed in Fig.~\ref{torque}. The torque per unit volume due to the demagnetizing field may be estimated as:

\begin {equation}
\Gamma_{demag} \approx 2 M^2 \Delta N \sin(4\theta)
\end{equation}

where $M$ is the magnetization and $\Delta N$ is the demagnetizing factor variation when the magnetic field is rotated in the plane. Using either the approximation of an ellipsoid, or the demagnetizing factor computed for a square-shaped sample\cite{Chen05}, we obtain $\Delta N \simeq$ 6 10$^{-2}$. Using typical value for the magnetization (e.g. $M \simeq$ 8 emu cm$^{-3}$ in the ordered state or $\chi \simeq$ 1 emu cm$^{-3}$ in the paramagnetic state at 280 K), we obtain $\Gamma_{demag} \simeq$ 10$^{-2}$ - 10$^{-1}$ Nm$^{-2}$, which is negligible, compared to the torque signal discussed in the rest.

\begin{figure}
\resizebox{0.6\columnwidth}{!}{%
  \includegraphics{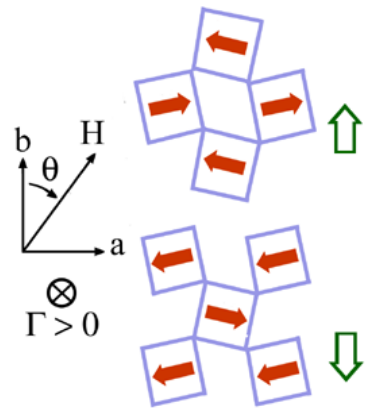}
  }
\caption{Left: conventions used for torque measurements. Right: two adjacent IrO$_2$ layers, antiferromagnetically coupled (torque is positive for the upper layer - open arrow is the net magnetization in each layer).}\label{sketchAF}
\end{figure}

\begin{figure}
\resizebox{0.9\columnwidth}{!}{%
  \includegraphics{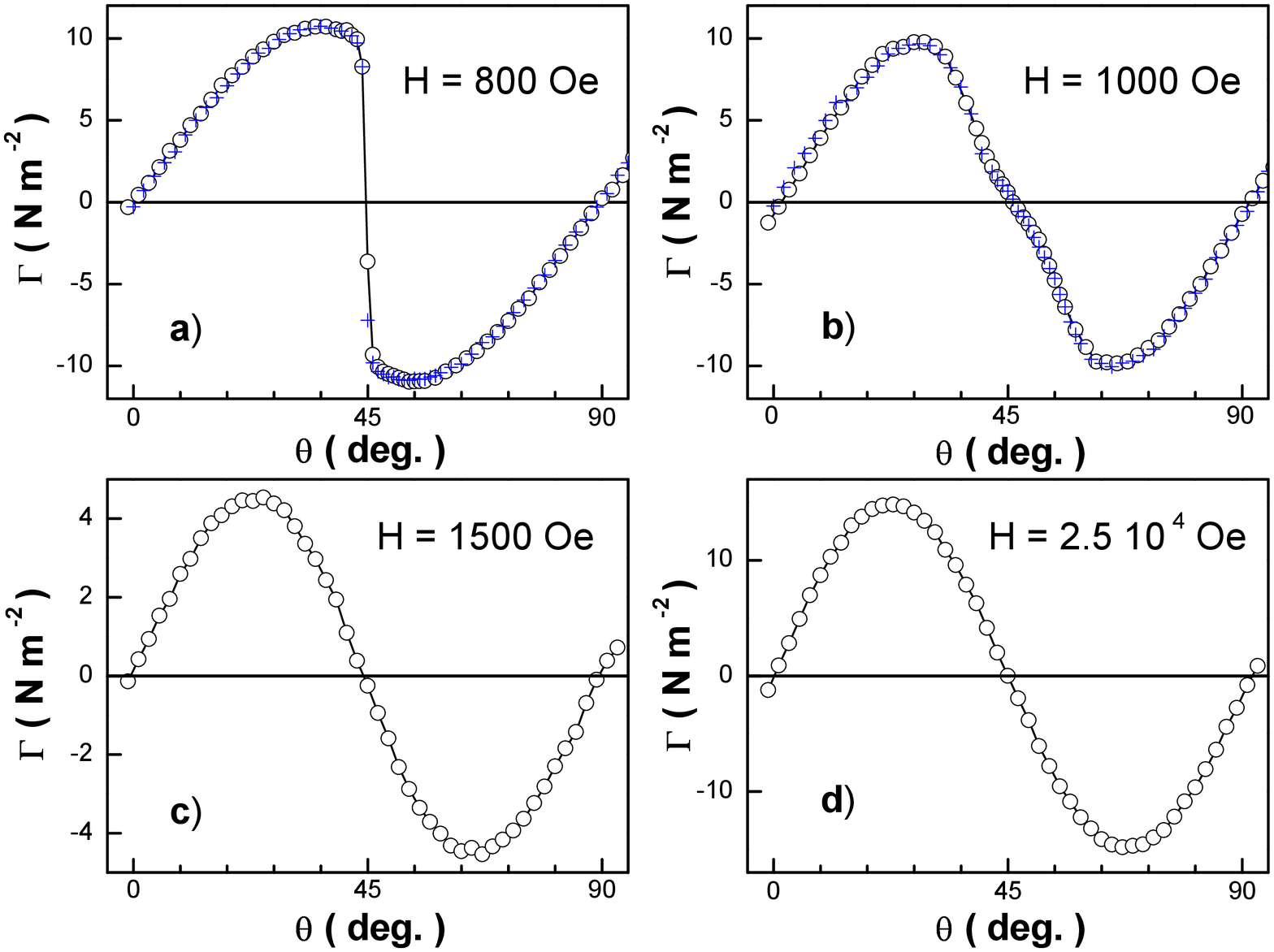}
  }
\caption{Torque for $T$ = 215 K. Circles are for increasing $\theta$; crosses, for decreasing ones. Note the difference in the torque scale. Conventions for $\theta$ and $\Gamma$ are displayed in Fig.~\ref{sketchAF}.}\label{torque}
\end{figure}

\begin{figure}
\resizebox{0.9\columnwidth}{!}{%
  \includegraphics{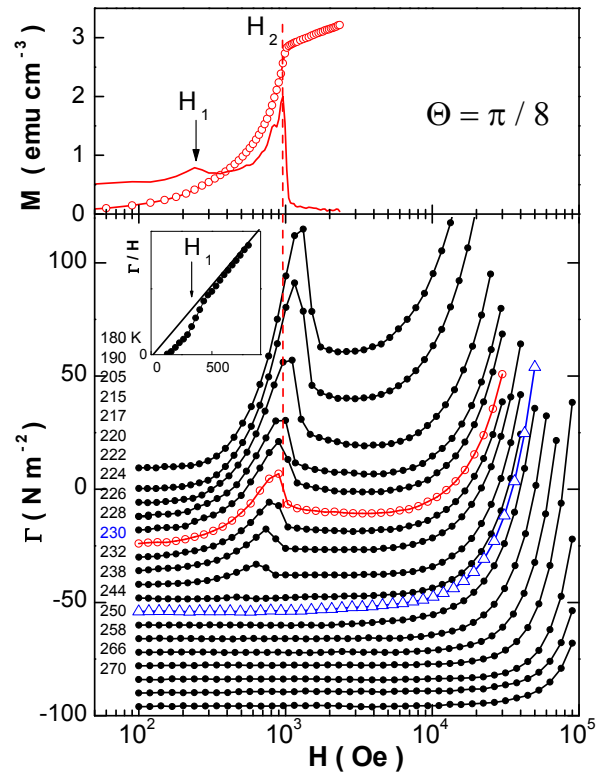}
  }
\caption{Upper: longitudinal magnetization ($T$ = 220 K) and it derivative, $dM/dH$. Lower: torque (curves have been shifted for clarity). The inset in the lower panel displays the transverse magnetization, $\Gamma / H$, at $T$ = 215 K, and triangles are for $T$ = 230 K.}\label{torque2}
\end{figure}

\begin{figure}
\resizebox{0.9\columnwidth}{!}{%
  \includegraphics{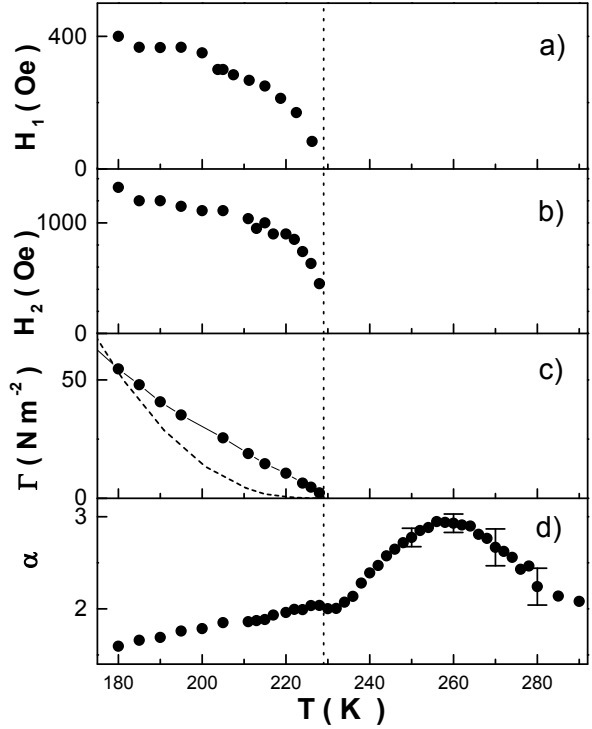}
  }
\caption{Quantities extracted from the data in Fig.~\ref{torque2}. a) First critical field, as given by both a step in the transverse and in the longitudinal magnetization. b) Field at the torque peak. c) Torque magnitude for $H$ = 0.4 T. The dotted line is $m(T)^{10}$ for the same field. d) Exponent for torque field dependence at high field, $\Gamma \propto H^\alpha$.}\label{torquefits}
\end{figure}

As expected from the quadratic symmetry of the crystal, the torque signal has a four-fold periodicity. Above some critical field $H_2 \simeq$ $10^3$ Oe, the signal is to a very good approximation sinusoidal (Fig.~\ref{torque}c,d), while, below this field, it shows a typical metastable behavior at $\theta$ = $\pi/4$  (Fig.~\ref{torque}a, b). Remarkably, the amplitude of the signal is not monotonous with the applied field, and a drop by about 50\% is observed at the crossing of $H_2$. A systematic study of the torque signal at a fixed angle confirms this feature, and allows to uncover some others. Figure \ref{torque2} shows the torque signal obtained for $\theta$ = $\pi/8$ for several temperatures, as well as the longitudinal magnetization at a selected temperature $T$ = 220 K. A sharp drop of the signal is observed at $H_2(T)$, also displayed in Fig.~\ref{torquefits}b. This coincides with the maximum slope of the longitudinal magnetization, $m(H)$ (Fig.~\ref{torque2}, upper panel). A much weaker feature is also present at a smaller field, $H_1(T)$. Upon crossing this field, the  longitudinal magnetization shows a small step (as evidenced by $dm/dH$ in the upper panel in Fig.~\ref{torque2}), and so does the transverse magnetization (Fig.~\ref{torque2}, inset in the lower panel). Both characteristic fields $H_1(T)$ and $H_2(T)$ sharply drop to zero at $T_c$ (Fig.~\ref{torquefits}).

For fields larger than $H_2(T)$, the torque signal shows a plateau where it is roughly independent of the applied field (this is most evident a few K from $T_c$, where the plateau is visible over a magnetic field decade). At still larger magnetic field, a noticeable increase of the torque signal occurs. It could be well fitted using a simple power law, $H^\alpha$, which we found could be valid over one field decade (the limitation being the maximum torque which our device could stand). Fitting the torque signal with such a power law, superimposed to a constant offset, yields the value of the plateau, as well as the exponent and the magnitude of the high field signal. It is found that the field-independent signal decreases to zero at $T_c$, in a quasi-linear way (Fig.~\ref{torquefits}c), while the high field exponent varies between $\alpha \simeq 1$ at low temperature and $\alpha \simeq 3$ at 260 K (Fig.~\ref{torquefits}d).

\subsection{Zero temperature model}

Clearly, the interpretation of the anisotropic magnetic properties of Sr$_2$IrO$_4$ is made difficult by the complex magnetic interactions hosted by this material. In the following, we introduce these interactions one after the other, in order to gauge the importance of the different contributions, and finally elaborate a model accounting for the observations.

First, we neglect the interlayer coupling and thus consider each layer separately. In this case, the magnetic configuration is essentially the one of a 2D antiferromagnet, with a four-fold anisotropy reflecting the quadratic symmetry of the crystal, with the additional feature of an in-plane ferromagnetic component (Fig.~\ref{sketchAF}). The torque for an anisotropic 2D antiferromagnet in an in-plane magnetic field was computed in ref.~\onlinecite{Herak10} and applied to the case of the cuprate Bi$_2$CuO$_4$. Essentially, the torque signal reflects the occurrence of a critical field, above which the AF domains flop to a configuration almost perpendicular to the applied field, which is well known as the spin-flop transition. Above this field, the AF domains nearly rigidly follow the applied magnetic field and the torque signal $\Gamma (\theta)$ is sinusoidal, with an amplitude independent of the applied field, $4 \, K_4$, where $K_4$ is the in-plane anisotropy constant in a phenomenological representation\cite{Herak10}. Identifying the intermediate-field torque plateau in Fig.~\ref{torque2} with this regime, the torque value in Fig.~\ref{torquefits}c is then simply $4 \, K_4 \sin(4\theta)$. This allows to estimate $K_4$ = 8 10$^3$ erg mol$^{-1}$ at $T$ = 180 K. It is then possible to estimate the spin-flop critical field, $H_{flop} = (16\,K/\,\chi_{\bot})^{1/2}$, where $\chi_\bot$ is the magnetic susceptibility for a magnetic field perpendicular to the spin-axis, and where we have neglected the susceptibility in the spin direction. Using the measured linear part of the magnetization $m(H)$ at $H$ = 7 T, we estimate $\chi_{\bot}$ = 6.5 10$^{-4}$ emu mole$^{-1}$ and, neglecting $\chi_\parallel$, $H_{flop}$ = 2 T at $T$ = 180 K. This is well above any of the characteristic fields evidenced by torque or longitudinal magnetization. 

Actually, the ferromagnetic component must be considered, as the driving torque on this component is larger than the one originating from the anisotropy of the susceptibility. Neglecting the anisotropic susceptibility contribution, it is easy to show that the critical field now is: $H_{flop} \simeq 4.3 K_4 /\, m_s$, where $m_s$ is the in-plane magnetization. Using the measured value $m_s$ = 470 emu mole$^{-1}$, one obtains $H_{flop}$ = 160 Oe at $T$ = 180 K. While this value can account for the lower field $H_1$, the model cannot account for the second field, $H_2$, as, for fields larger than $H_1$, one merely expects a rigid rotation of the layers. 

To account for this second field, one needs to introduce the AF coupling between layers. The simplest model to do so is to introduce the ferromagnetic components in two adjacent layers, and a phenomenological expression for the magnetic total energy as the sum of two terms:

\begin{eqnarray}
F_s  =  \sum_{l=1,2} -K_4 \cos(4\arccos(\textbf{m}_{l} \cdot \textbf{a})) - \textbf{m}_{l} \cdot \textbf{B} \nonumber \\
F_i  = J_\bot \, \textbf{m}_{1} \cdot \textbf{m}_{2}
\label{eq1}
\end{eqnarray}

where the sum is for two adjacent layers, $l$ = 1 and $l$ = 2; $\textbf{m}_l$ is the ferromagnetic component in layer $l$, \textsl{$F_{s}$} contains the magnetocrystalline and Zeeman energies, and \textsl{$F_{i}$} is for the antiferromagnetic coupling of the two layers. Although this model applies to two distinct layers, it is essentially the same as the one in Ref.~\onlinecite{Herak10} (using $\chi_\bot \sim 1/J$), with the difference that -- as the magnetic field may now reach values corresponding to the flip of the AF-coupled magnetizations -- one cannot longer treat these within the anisotropic susceptibility approximation . As a result, one expects from Eq.~\ref{eq1} a critical field for the flop of the ferromagnetic domains similar to the one estimated above, which accounts for $H_1$, and a critical field for the flip of the AF-coupled magnetizations towards the parallel configuration, which accounts for $H_2$. Thus, this model offers a possibility to estimate two credible critical fields, but it still cannot account completely for the observed torque signal. In particular, it only predicts a saturation of the torque at $H_2$, in place of the observed non-monotonic behavior. 

We found that a realistic model may only be obtained by taking into account both the spin in-plane degrees of freedom, as for the first model, and the out-of-plane coupling, as for the second one. The total energy is then the sum of three terms:

\begin{eqnarray}
F_s = \sum_{l=1,2 \,i=1,2} -K_4 \cos(4\arccos(\textbf{S}^{i}_{l} \cdot \textbf{a})) - \textbf{S}^{i}_{l} \cdot \textbf{B} \nonumber \\
F_a = \sum_{l=1,2} J\, \textbf{S}^{1}_{l} \cdot \textbf{S}^{2}_{l} - D\, \textbf{S}^{1}_{l} \wedge \textbf{S}^{2}_{l} \nonumber \\
F_i = J_\bot \,(\textbf{S}^{1}_{1}+\textbf{S}^{2}_{l}) \cdot (\textbf{S}^{1}_{2}+\textbf{S}^{2}_{2})
\label{eq2}
\end{eqnarray}

where $l = 1,2$ is for the two coupled layers; $i = 1,2$ is for the two spins of one pair; $J$ is the antiferromagnetic coupling of two spins belonging to the same plane; $D$ is the Dzyalochinskii-Morya coefficient (which drives the tilt of the spins, and so generates the ferromagnetic in-plane component \textbf{m$_l$} as in Eqs.~\ref{eq1}), and $J_\bot$ antiferromagnetically couples \textbf{m$_1$} and \textbf{m$_2$}. Strictly, the first term should be $-K_4 \cos(4(\arccos(\textbf{S}^{i}_{l} \cdot \textbf{a}))-\alpha)$, where $\alpha \simeq D/2J$, to account for the octahedra rotation, but this is irrelevant for the present simulations.

\begin{figure}
\resizebox{0.9\columnwidth}{!}{%
  \includegraphics{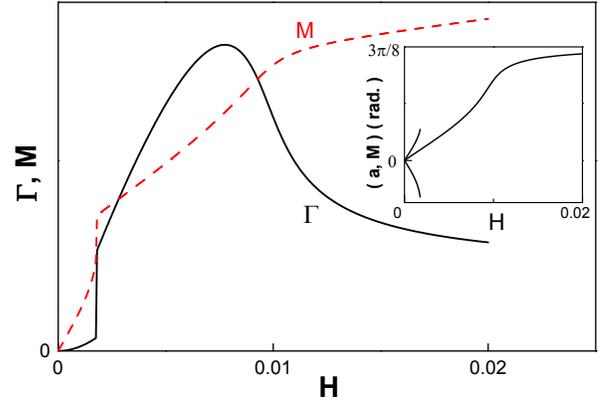}
  }
\caption{Torque and longitudinal magnetization, as obtained from the model of Eqs.~\ref{eq2}. The inset displays the angle of the magnetization in the two layers, from the \textit{a}-axis. Parameters for the simulation are $J$ = 1, $S$ = 1, $K_4$ = 6 10$^{-5}$, $J_\bot$ = 6 10$^{-4}$ and $D$ = 0.1. The magnetic field is applied at $\theta$ = $\pi/8$.}\label{simulation}
\end{figure}

As may be seen in Fig.~\ref{simulation}, the model reproduces the essential experimental features for both the longitudinal magnetization and the torque. It is seen that, below $H_1$, two domain orientations coexist while, above $H_2$, the magnetizations in the two adjacent layers are ordered ferromagnetically and progressively rotate towards the magnetic field orientation, as the field is increased. The peak in the torque magnitude does not mark a transition delimiting two distinct spin configurations, but a crossover where the spin canting is found to vary by about 10\%. While the only true transition at $H_1$ produces a well marked jump in the torque and magnetization simulation, we have seen (Fig.~\ref{torque2}) that the experimental manifestation is actually weak. This may be explained both by a spatial distribution of the material properties, and by the fact that the simulation postulates identical populations of the two domain orientation below $H_1$, and a single stable domain above this field, while pinning and inhmogenities may well smear out the singularity.

\subsection{Basal-plane anisotropy}

Within the modelization made above, the observation, below $T_c$, of a torque signal increasing with field at large field should be interpreted as the manifestation of a field-dependent parameter $K_4$. Van Vleck first laid the basis for a quantum theory for the computation of the magnetocrystalline anisotropy in ferromagnetic materials\cite{Vanvleck37}. He showed that the anisotropy parameter may be evaluated using an effective anisotropic spin-Hamiltonian in the local Weiss field approximation (as a result of crystal field splitting and spin-orbit coupling), and the statistical computation of the spin orientation distribution derived from this Hamiltonian. This distribution also determines the magnetization, and, thus, the anisotropy may be expressed as a function of this measurable quantity (see Ref.~\cite{Callen66} for a review). At low temperature, when spin-waves do not destroy the two-spin correlation, this yields the well-known power law dependence, $\kappa_l \propto$ $M^{l(l+1)/2}$, where $\kappa_l$ are the coefficients of the magnetic energy in a spherical harmonics representation\cite{Zener54}. The effective anisotropy is then proportional to one of these coefficients, or a linear combination of them. This relationship does not depend on the details of the spin interactions, but only on the symmetry of the crystal, and the order of the anisotropy. It was shown that the mean-field approximation is actually an example of a renormalized collective excitation theory, to which also belongs the spin-wave theory\cite{Callen65}. In this hypothesis that the spin excitations are quasi-independent, the two limiting behaviors $\kappa_l \propto M^{l(l+1)/2}$ and $\kappa_l \propto M^l$, respectively in the ordered and the paramagnetic states, are predicted. In the case of the tetragonal symmetry, the correspondence between these coefficients and the basal plane four-fold anisotropy constant is straightforward, as $K_4 \propto \kappa_4$.

The interpretation of the power law field dependence for $K_4$ then follows for each temperature regime. At low temperature, $M(H)$ is the sum of the field-independent $M_s$ and a small linear field term, so that any power law of the magnetization yields $\alpha \simeq$ 1. Indeed, $\alpha$ decreases steadily with decreasing temperature in Fig.~\ref{torquefits}d, reaching $\alpha = 1.4$ at $T$ = 150 K. At $T_c$, the critical scaling gives $M \propto H^{\beta/(\beta+\gamma) \simeq 0.2}$ (Section~\ref{LongMag}), so that $K_4 \propto M^{10} \propto H^2$, as observed. Finally, in the paramagnetic regime, one expects $K_4 \propto M^4 \propto H^4$. While the measured exponent grows up to $\alpha \simeq 3$ at $T$ = 260 K, it is however seen to decrease for larger temperature, down to $\alpha \simeq 2$. This could be due to the fact that the contribution of the anisotropy becomes quickly smaller for higher temperature (as $M(T)^4$), and the torque signal eventually becomes dominated by some anisotropic contribution of the form $\Delta \chi H^2$ (i.e. $\propto$ $M^2$). It is found, however, that $M(T)^{10}$ at low magnetic field does not account for the quasi-linear temperature dependence for $K_4(T)$ when $T \rightarrow T_c$ (Fig.~\ref{torquefits}c). This discrepancy could sign the breakdown of the two-spins correlations by thermal spin-waves, when $k_B T$ becomes larger than the typical spin-wave energy for wave vector $k \simeq a^{-1}$. Still, we evaluate the zero temperature anisotropy from the extrapolation of $K_4$ at 150 K to $K_4(0)$ = 1.2 10$^5$ erg mole$^{-1}$, using the $M^{10}$ scaling. This value is one order of magnitude larger than the one of Bi$_2$CuO$_4$, which should be representative of tetragonal cuprates\cite{Herak10}.

In Ref.~\onlinecite{Vale15}, it was proposed that the anisotropy is large enough to influence the universality class of the fluctuations. In a classical description, it is not expected to do so, as long as the four-fold contribution to the free energy is smaller than the spin coupling energy, $J$. More precisely, Ref.~\onlinecite{Taroni08} determines $K_4/J \simeq$ 0.5. The anisotropy energy, $\approx$ 0.13 $\mu$eV/spin, is only about 10$^{-6} \,J$. The observation was made, however, that 2D quantum confinement yields a larger effective anisotropy for planar antiferromagnets, as a small anisotropy term opens a large magnon gap on the scale of $J$\cite{Vale15}. It was proposed that the effective anisotropy is in this case $(24 K_4\,J)^{1/2}$ $\simeq$ 6 10$^{-3} J$. This modified value is smaller than the upper bound obtained from the magnon dispersion in Ref.~\onlinecite{Vale15}, 8 $10^{-2}\,J$; it is also too small to expect a noticeable deviation from the isotropic 2D XY model.

Finally, we show that torque measurements also provide a way to bound possible anisotropic thermodynamic effects. In a previous contribution, we proposed that an increase of the ordering temperature with field could be at the origin of the observation of magnetoresistance effects above $T_c$\cite{Fruchter15}. In particular, we recalled that the Dzyalochinskii-Morya term is the origin of a field-induced, transverse, staggered magnetic field ($H^\dag$), and, so, of a transverse staggered magnetization. This staggered field competes with the conventional suppression of $T_c$ with field and, at low field, may increase the ordering temperature in a linear way\cite{Kagawa08}. The simplest hypothesis to evaluate this effect above $T_c$ is to assume a shift of the Curie-Weiss temperature, as it should most directly reflect the change in the local spin-coupling with the induced staggered field. The shift due to the staggered field may be estimated as $\Delta T_{CW}/T_{CW} = H^\dag \mu / J$, where $H^\dag = z \left\langle M\right\rangle D / \mu^2$, $z = 4$ is the coordination of the magnetic lattice, $\left\langle M\right\rangle$ the average magnetization per spin at $T \approx T_{CW}$, and $\mu$ the moment of the Ir atom. $H^{-1} \Delta T_{CW}/T_{CW}$ may be evaluated in this way as large as 3 10$^{-3}$ T$^{-1}$. The anisotropic part of this shift is however not known (it was erroneously assumed that there should be one in Ref.~\cite{Fruchter15}, but the absence of a projection of the Dzyalochinskii-Morya vector on the basal plane does not allow for this direct source of anisotropy, in the present case, unlike for La$_2$CuO$_4$). Assuming a four-fold variation $T_{CW} = \Delta_4 T_{CW} \sin(4\theta)$, the associated contribution to torque is $4 B M \Delta_4 T_{CW} / (T-T_{CW})$. Using the torque amplitude measured at 260 K and 6 T as the maximum contribution for this effect, we obtain $H^{-1} \Delta_4 T_{CW}/T_{CW} <$ 10$^{-5}$ T$^{-1}$. So, the anisotropy of the ordering temperature shift must be very small, if any.

\section*{Conclusion}

To summarize, we have shown that the bulk magnetization critical scaling across $T_c$ is close to the one expected for 2D XY scaling, as was found earlier from the temperature dependence of the magnetic coherence length, above $T_c$. There is no observable effect of an in-plane magnetic field on the fluctuations dimensionality, and the scaling as $T \rightarrow$ 0 indicates the possible importance of three-dimensional fluctuations in this limit.
We have modeled the longitudinal and transverse magnetizations, taking into account the basal-plane couplings and anisotropy, as well as the transverse coupling. We find that the basal-plane anisotropy is too small to account for large deviations from an isotropic 2D XY model, using a simple estimate for the effective anisotropy enhancement from quantum fluctuations effects.

\section*{}

L.F. performed the experiments and wrote the paper, with inputs from co-authors, who also provided samples.
We acknowledge support from the Agence Nationale de la Recherche grant SOCRATE.

%
%
%
%
%

\end{document}